\documentclass[5p,times,preprint]{elsarticle}
\usepackage{lineno,hyperref}
\usepackage{amsfonts}
\usepackage{amsmath}
\usepackage{graphicx}
\modulolinenumbers[5]
\usepackage{booktabs} 
\usepackage{float} 
\usepackage{enumitem}
\usepackage{numprint}
\usepackage{newtxtext}
\usepackage[varvw]{newtxmath}
\usepackage{balance}
\usepackage{textcomp}
\journal{Nuclear Engineering and Technology}

\begin{document}

\begin{frontmatter}
		
\title{Calculation of kinetic parameters $\beta_{\mathit{eff}}$ and $\Lambda$ with modified open source Monte Carlo code OpenMC(TD)}

\author[add1,add2]{J. Romero-Barrientos}
\ead{romeroj@uchile.cl}
\cortext[mycorrespondingauthor]{Corresponding Author}
\author[add3]{J.I. M\'arquez Dami\'an}
\author[add1,add5]{F. Molina}
\author[add1,add6]{M.~Zambra}
\author[add1,add4]{P.~Aguilera}
\author[add1,add2]{F.~L\'opez-Usquiano}
\author[add7]{B.~Parra}
\author[add1,add4]{A.~Ruiz}

\address[add1]{Comisi\'on Chilena de Energ\'ia Nuclear, Nueva Bilbao 12501, Las Condes, Santiago, Chile}
\address[add2]{Departamento de F\'isica, Facultad de Ciencias F\'isicas y Matem\'aticas, Universidad de Chile, Blanco Encalada 2008, Santiago, Chile}
\address[add3]{Spallation Physics Group, European Spallation Source ERIC, P.O. Box 176, 22100 Lund, Sweden}
\address[add4]{Facultad de Ciencias, Departamento de F\'isica, Universidad de Chile, Las Palmeras 3425, \~Nu\~noa, Santiago,Chile}
\address[add5]{Departamento de Ciencias Físicas, Universidad Andres Bello, Sazi\'e 2212, 837-0136, Santiago, Chile}
\address[add6]{Universidad Diego Portales, Manuel Rodr\'iguez Sur 415, Santiago, Chile}
\address[add7] {Instituto de F\'isica Corpuscular, Parque Cient\'ifico, C/Catedr\'atico Jos\'e Beltr\'an, 2, E-46980 Paterna, España}



\begin{abstract}
This work presents the methodology used to expand the capabilities of the Monte Carlo code OpenMC for the calculation of reactor kinetic parameters: effective delayed neutron fraction $\beta_{\mathit{eff}}$ and neutron generation time $\Lambda$. The modified code, OpenMC(Time-Dependent) or OpenMC(TD), was then used to calculate the effective delayed neutron fraction by using the prompt method, while the neutron generation time was estimated using the pulsed method, fitting $\Lambda$ to the decay of the neutron population. OpenMC(TD) is intended to serve as an alternative for the estimation of kinetic parameters when licensed codes are not available. The results obtained are compared to experimental data and MCNP calculated values for $18$ benchmark configurations.  
\end{abstract}

\begin{keyword}
OpenMC, Monte Carlo, kinetic parameters, open source, neutron generation time, effective delayed neutron fraction 
\end{keyword}

\end{frontmatter}

\section{Introduction}
The determination of the effective delayed neutron fraction $\beta_{\mathit{eff}}$, and the mean neutron generation time $\Lambda$, play a key role in reactor kinetics. $\beta_{\mathit{eff}}$ and $\Lambda$ values indicate the safe reactor operation limits for a given fuel, and the prompt neutron lifetime depending on the fuel burnup, respectively. The estimation of these quantities requires calculating the adjoint neutron flux or taking into account the system time-dependence. One way to achieve the former is by using Monte Carlo transport code MCNP~\cite{Goorley2012}, while the latter requires the time monitoring of neutron population. 

\begin{sloppypar}
In this work the open source Monte Carlo code OpenMC~\cite{Romano2015} developed at the Massachusetts Institute of Technology was modified to explicitly include time-dependency, so kinetic parameters can be estimated. The modified code was named Time-Dependent OpenMC or OpenMC(TD).
The code was tested against nineteen benchmarks, where the models were taken from the ICSBEP {\it International Handbook of Evaluated Criticality Safety Benchmark Experiments}~\cite{icsbep}, while the benchmark results were taken from the literature~\cite{leppanen}. These benchmarks range from the Godiva bare sphere system to the RA-$6$ open pool research reactor, comparing the measured kinetic parameters with calculated values using MCNP and OpenMC(TD). The aim of this work is to show an alternative way to obtain nuclear reactor kinetic parameter values using an open source Monte Carlo code.
\end{sloppypar}

\section{OpenMC Monte Carlo code}
\label{section:openmc}
\begin{sloppypar}
The Monte Carlo code OpenMC~\cite{Romano2015} is a relatively new, open-source code for particle transport developed by the Massachusetts Institute of Technology since $2013$. This code is capable of simulating neutron transport in fixed source, $k$-eigenvalue, and subcritical multiplication problems. The code supports both continuous-energy and multigroup cross sections. OpenMC estimates physical quantities, such as neutron flux, through \textit{tallies}; \textit{filters} are used to specify a region of the phase space, for example, scoring neutron flux for a given cell position. The continuous-energy nuclear cross section data follows the HDF5 format~\cite{Koranne2011}  and is generated from ACE files produced by NJOY~\cite{osti_1338791}.
This code is open source, and its license allows to modify the source code to develop and add new capabilities. An example of this last point is the many number of publications associated to the code, covering topics and developments such as benchmarking~\cite{khurrum}, coupling and multi-physics~\cite{CHEN2017264}, geometry and visualization~\cite{LI2018329}, multigroup cross section generation~\cite{variansyah}, doppler broadening~\cite{walsh}, nuclear data~\cite{walsh_j}, parallelism~\cite{forget}, depletion~\cite{binhang} and sensitivity analysis~\cite{xingjie}.
\end{sloppypar}
\section{Methodology}
\label{section:methods}
\subsection{Addition of time-dependent capabilities to OpenMC}

In a stationary Monte Carlo transport simulation, time is not explicitly present and for the same reason, quantities that change in time cannot be directly obtained. Consequently, the first step to include time explicitly in a Monte Carlo simulation was adding a new label $t$ to the particles, serving as a \textit{clock}, which value was updated using the kinetic energy and the distance traveled by the neutron between any interaction. This time was set to zero ($t\!=\!0$) at the beginning of the simulation, i.e. when the first particle is emitted from the source, and it was updated while the particle was transported through the geometry of the simulation.
After adding the time label to the simulation and in order to account the measured quantities in time, the OpenMC \textit{tallies} were modified and a new \textit{time filter} was included. This filter added the capability to monitor the time evolution of any of the tallies already present in the code. This capability allows to tally, for example, neutron flux as a function of time. The user can define the length and number of the time intervals as desired. This was implemented as part of the PhD thesis of J. Romero-Barrientos~\cite{arxiv-moi}, using the version of OpenMC $0.10$, written in Fortran. The current version of OpenMC is $0.12$, written in C++.

\subsection{Effective delayed neutron fraction $\beta_{\mathit{eff}}$}

The effective delayed neutron fraction is defined as the ratio between the adjoint neutron flux and the spectrum weighted number of fissions induced by delayed neutrons and number of fissions induced by all neutrons, 

\begin{equation}
\beta_{\mathit{eff}} = \frac{\int \phi^{\dagger} (\vec{r},E,\vec{\Omega}) \chi_d(E) \beta \nu(E) \Sigma_f(\vec{r},E) \phi(\vec{r},E,\vec{\Omega}) dE d\vec{\Omega} d\vec{r}}{\int \phi^{\dagger} (\vec{r},E,\vec{\Omega}) \chi(E) \nu(E) \Sigma_f(\vec{r},E) \phi (\vec{r},E,\vec{\Omega}) dE d\vec{\Omega} d\vec{r}},
\label{eq:betaeff_formal_definition}
\end{equation}
where $\phi$ is the neutron flux, $\phi^{\dagger}$ is the adjoint neutron flux, $\vec{r}$ is the neutron position, $E$ is the neutron energy, $\vec{\Omega}$ is the neutron direction, $\chi_d(E)$ the delayed neutron energy spectrum, $\chi(E)$ the fission neutron energy spectrum, $\beta$ the delayed neutron fraction, $\nu(E)$ is the average number of neutrons released per fission, and $\Sigma_f$ is the macroscopic fission cross section. 
\begin{sloppypar}
MCNP can calculate the effective delayed neutron fraction, Rossi-$\alpha$ and neutron generation time values using the iterated fission probability method, computing adjoint-weighted tallies in criticality calculations~\cite{LA-UR-10-01700}.
\end{sloppypar}
As it can be seen in Eq.~\eqref{eq:betaeff_formal_definition}, formally the adjoint neutron flux is needed to estimate $\beta_{\mathit{eff}}$, but in this work, the effective delayed neutron fraction was calculated using the prompt method~\cite{Meulekamp2006}. This method estimates the value of $\beta_{\mathit{eff}}$ by: 

\begin{equation}
\beta_{\mathit{eff}} \sim 1 - \frac{k_p}{k_{\mathit{eff}}},
\label{eq:beta_eff}
\end{equation}
where $k_{\mathit{eff}}$ is the effective multiplication factor obtained from a criticality calculation, and $k_{\mathit{p}}$ is the effective multiplication factor obtained without taking into account the delayed neutrons contribution. With the prompt method the adjoint neutron flux is not needed to calculate $\beta_{\mathit{eff}}$. Thus, criticality calculation without delayed neutrons was added as a new capability to OpenMC(TD). This enabled the estimation of $\beta_{\mathit{eff}}$ by running a criticality calculation for $k_{\mathit{eff}}$ and other for $k_p$.

\subsection{Mean neutron generation time $\Lambda$}
Another important kinetic parameter to characterize the dynamics of a nuclear reactor is the mean neutron generation time, $\Lambda$, defined as
\begin{equation}
\label{eq:Lambda_general}
\Lambda = \frac{l}{k},
\end{equation}
where $l$ is the prompt neutron lifetime and $k$ is the multiplication factor. This means that $\Lambda$ is the time between the birth of a neutron and the subsequent absorption that induces fission. Formally, the neutron generation time is defined in terms of the neutron flux and adjoint neutron flux:

\begin{equation}
\Lambda = \frac{\int \phi^{\dagger} (\vec{r},E,\vec{\Omega}) \frac{1}{\varv(E)} \phi (\vec{r},E,\vec{\Omega}) dE d\vec{\Omega} d\vec{r}}{\int \phi^{\dagger} (\vec{r},E,\vec{\Omega}) \chi(E) \nu(E) \Sigma_f(\vec{r},E) \phi (\vec{r},E,\vec{\Omega}) dE d\vec{\Omega} d\vec{r}},	
\label{eq:Lambda_integral}
\end{equation}
where $\varv(E)$ is the neutron speed. 

In this work the neutron generation time was calculated using the pulsed method~\cite{simmons, snoj}, where a neutron pulse is inserted into a subcritical system and then the neutron population decay is measured. A theoretical approximation used to study this decay is the point kinetics approximation, where the flux is assumed to be a separable function of space and time~\cite{bell}. Under this assumption the neutron population differential rate, $dn/dt$, is given by

\begin{equation}
	\frac{dn}{dt} = \frac{\rho - \beta_{\mathit{eff}}}{\Lambda} n(t) + \sum_j \lambda_j C_j(t),
	\label{eq:dn/dt}
\end{equation}
where $\rho$ is the system reactivity, $\lambda_j$ and $C_j$ are the decay constant and concentration for the $j$-th precursor group, respectively. Since the phenomena being studied is the prompt neutron decay, the contribution from delayed neutrons can be neglected, 
\begin{equation}
\frac{\partial}{\partial t} n(t) = \frac{\rho - \beta_{\mathit{eff}}}{\Lambda} \,n(t).
\label{eq:neutrons_pulsed}
\end{equation} 
The solution to Eq.~\eqref{eq:neutrons_pulsed} is given by
\begin{equation}
n(t) = n_0 e^{\alpha t}, \quad \text{with} \quad \alpha= \frac{\rho - \beta_{\mathit{eff}}}{\Lambda}
\label{eq:prompt_decay}
\end{equation}
where $n_0$ is the initial neutron population and $\alpha$ is the decay constant. 
The expanded time-dependent capabilities added to OpenMC(TD), described in Sec.~\ref{section:methods}, were used to obtain the decay of the neutron flux as a function of time after placing a neutron source in a subcritical system. The reactivity of the system, $\rho$, was calculated in a previous criticality run, and $\beta_{\mathit{eff}}$ was estimated by using the prompt method. Finally, $\Lambda$ was obtained by fitting Eq.~\eqref{eq:prompt_decay} to the neutron population decay curve from the simulation.

\section{OpenMC(TD) benchmarks}
\label{section:benchmark}
The modifications added to the OpenMC code were tested using a set of nineteen benchmarks taken from the International Handbook of Evaluated Criticality Safety Benchmark Experiments~\cite{icsbep}. Different cases studied in this work includes uranium, plutonium, and U-$233$ fuel systems, with both thermal and fast neutron energy spectra:

\begin{itemize}[label=\raisebox{0.25ex}{\tiny$\bullet$}]
	\setlength\itemsep{0.5pt}
	\item \textit{Godiva (HEU-MET-FAST-001)}: Bare sphere of highly enriched uranium. The isotopes present in this benchmark are ${}^{233}$U, ${}^{235}$U and ${}^{238}$U. 
	\item \textit{Jezebel-Pu (U233-MET-FAST-001)}: Bare sphere of plutonium. The isotopes present in this benchmark are ${}^{\mathit{nat}}$Ga, ${}^{239}$Pu, ${}^{240}$Pu and ${}^{241}$Pu.
	\item \textit{Skidoo (U233-MET-FAST-001)}: Bare sphere of uranium, were ${}^{233}$U is at $98 \%$. The isotopes present in this benchmark are ${}^{233}$U, ${}^{234}$U, ${}^{235}$U and ${}^{238}$U.
	\item \textit{Topsy (HEU-MET-FAST-028)}: Enriched uranium sphere surrounded by a reflector of natural uranium. The isotopes present in this benchmark are ${}^{234}$U, ${}^{235}$U and ${}^{238}$U.
	\item \textit{Popsy (PU-MET-FAST-006)}: Plutonium sphere surrounded by a reflector of natural uranium. The isotopes present in this benchmark are ${}^{\mathit{nat}}$Ga, ${}^{239}$Pu, ${}^{240}$Pu and ${}^{241}$Pu,  ${}^{234}$U, ${}^{235}$U and ${}^{238}$U.
	\item \textit{Flattop$23$ (U233-MET-FAST-006)}: Uranium sphere surrounded by a reflector of natural uranium. The isotopes present in this benchmark are ${}^{233}$U, ${}^{234}$U, ${}^{235}$U and ${}^{238}$U.
	\item \textit{BigTen (IEU-MET-FAST-007)}: Mixed uranium-metal cylindrical core with $10 \%$ average ${}^{235}$U enrichment surrounded by a reflector of depleted uranium. The isotopes present in this benchmark are ${}^{234}$U, ${}^{235}$U, ${}^{236}$U, and ${}^{238}$U.
	\item \textit{ZPR-U$9$ (IEU-MET-FAST-010)}: Cylindrical assembly of uranium metal with a depleted uranium reflector. The isotopes present in this benchmark are ${}^{234}$U, ${}^{235}$U, ${}^{236}$U, ${}^{238}$U, Ni, Cr, Fe, Si, Mo, Cu, ${}^{27}$Al, C, Mg, ${}^{1}$H, ${}^{2}$H, and ${}^{19}$F.
	\item \textit{SNEAK-$7$A, -$7$B}: Unmoderated PuO${}_{2}$/UO${}_{2}$ with a depleted uranium reflector. The isotopes present in this benchmark are ${}^{235}$U, ${}^{238}$U, ${}^{239}$Pu, ${}^{240}$Pu, ${}^{241}$Pu, ${}^{242}$Pu, Ni, Cr, Fe, Si, Mo, ${}^{27}$Al, C, Mg, and O.  
	\item \textit{TCA}: Light water, moderated and low-enriched UO${}_{2}$ core in a tank-tpe critical assembly. The isotopes present in this benchmark are ${}^{234}$U, ${}^{235}$U, ${}^{238}$U, ${}^{1}$H, ${}^{16}$O, and ${}^{27}$Al.              
	\item \textit{Stacy-$029$, -$033$, -$046$ (LEU-SOL-THERM-004)}: Water-reflected cylindrical tank with uranyl-nitrate solution at $10 \%$ enrichment. The isotopes present in this benchmark are ${}^{234}$U, ${}^{235}$U, ${}^{238}$U, ${}^{1}$H, ${}^{14}$N, and ${}^{16}$O.      
	\item \textit{Stacy-$030$ (LEU-SOL-THERM-007)}: Water-reflected cylindrical tank with uranyl-nitrate solution at $10 \%$ enrichment. The isotopes present in this benchmark are ${}^{234}$U, ${}^{235}$U, ${}^{238}$U, ${}^{1}$H, ${}^{14}$N, and ${}^{16}$O.
	\item \textit{Stacy-$125$ (LEU-SOL-THERM-016)}: Water-reflected slabs of $10 \%$ enriched uranyl nitrate solution. The isotopes present in this benchmark are ${}^{234}$U, ${}^{235}$U, ${}^{236}$U, ${}^{238}$U, ${}^{1}$H, ${}^{16}$O, ${}^{17}$O, C, Si, Ni, Fe, and Cr.
	\item \textit{RA-$6$ (IEU-COMP-THERM-014)}: Open pool research reactor, fuel elements are MTR-type with $19.7 \%$ enrichment. Two configurations were studied, which are detailed in Sec.~\ref{subsection:RA6}
\end{itemize}
The systems described above are summarized in Table~\ref{table:systems}. 
\begin{table*}[h!]
	\centering
	\footnotesize
	\catcode`?=\active \def?{\phantom{0}}
	\catcode`!=\active \def!{\phantom{i}}
	{\def\arraystretch{1.2}\tabcolsep=20pt
	\begin{tabular}{@{}llll@{}}
		\toprule
		Case & Case                      & Case   & Benchmark \\ 
		number & name                       & description   & data \\ \midrule
		$1$ & Godiva                       & Bare and homogeneous sphere of highly enriched uranium.   & $k_\mathit{eff}$, $\beta_{\mathit{eff}}$ \\
		$2$ & Jezebel-Pu                   & Bare and homogeneous sphere of plutonium. & $k_\mathit{eff}$, $\beta_{\mathit{eff}}$ \\
		$3$ & Skidoo                       & Bare sphere of U-$233$. & $k_\mathit{eff}$, $\beta_{\mathit{eff}}$ \\
		$4$ & Topsy                        & Highly enriched uranium sphere surrounded by a natural uranium reflector. & $k_\mathit{eff}$, $\beta_{\mathit{eff}}$ \\
		$5$ & Popsy                        & Plutonium sphere surrounded by a natural uranium reflector. & $k_\mathit{eff}$, $\beta_{\mathit{eff}}$ \\
		$6$ & Flattop$23$                  & U-$233$ sphere surrounded by a natural uranium reflector. & $k_\mathit{eff}$, $\beta_{\mathit{eff}}$ \\
		$7$ & BigTen                       & Intermediate enriched uranium cylindrical core surrounded by a natural uranium reflector. & $k_\mathit{eff}$, $\beta_{\mathit{eff}}$ \\
		$8$ & ZPR-U$9$                     & Cylindrical assembly of uranium metal with a depleted uranium reflector. & $k_\mathit{eff}$, $\beta_{\mathit{eff}}$ \\
		$9$ & SNEAK-$7$A                  & Unmoderated PuO${}_{2}$/UO${}_{2}$ core with a depleted uranium reflector. & $k_\mathit{eff}$, $\beta_{\mathit{eff}}$ \\
		$10$ & SNEAK-$7$B                  & Unmoderated PuO${}_{2}$/UO${}_{2}$ core with a depleted uranium reflector. & $k_\mathit{eff}$, $\beta_{\mathit{eff}}$ \\
		$11$ & TCA                         & Light water moderated low enriched UO${}_{2}$ core in tank-type critical assembly. & $k_\mathit{eff}$, $\beta_{\mathit{eff}}$ \\
		$12$ & Stacy-$029$                 & Water-reflected cylindrical tank with uranyl nitrate solution. & $k_\mathit{eff}$, $\alpha_R$ \\
		$13$ & Stacy-$033$                 & Water-reflected cylindrical tank with uranyl nitrate solution. & $k_\mathit{eff}$, $\alpha_R$ \\
		$14$ & Stacy-$046$                 & Water-reflected cylindrical tank with uranyl nitrate solution. & $k_\mathit{eff}$, $\alpha_R$ \\
		$15$ & Stacy-$030$                 & Unreflected cylindrical tank with uranyl nitrate solution. & $k_\mathit{eff}$, $\alpha_R$ \\
		$16$ & Stacy-$125$                 & Water reflected slabs of enriched uranyl nitrate solution .& $k_\mathit{eff}$, $\alpha_R$ \\
		$17$ & RA-$6$ (Case $1$)  & Intermediate enriched uranium open pool research reactor. Water reflected configuration. & $k_\mathit{eff}$ \\ 
		$18$ & RA-$6$ (Case $2$) & Intermediate enriched uranium open pool research reactor. Graphite reflected configuration. & $k_\mathit{eff}$, $\alpha_R$ \\ \bottomrule
	\end{tabular}
    }
	\caption{Description of benchmarks studied in this work.}
	\label{table:systems}
\end{table*}

\subsection{Description of the RA-6 reactor}
\label{subsection:RA6}
The RA-6 (spanish acronym for Argentina Reactor, Number 6) is an open pool research reactor with a nominal power of $3$~MW, located in the Bariloche Atomic Center, San Carlos de Bariloche, Río Negro, Argentina.
The core of the reactor is made up of an array of flat plate MTR-type fuel elements with $20 \%$ enriched uranium located inside a stainless steel tank filled with demineralized water that acts as a coolant, moderator, reflector and shielding in the axial direction. Reactivity is controlled by four Ag--In--Cd plates. The fuel elements were modeled explicitly, including the cadmium wires, water gaps, guides and nozzles. The model also included the supporting grid for the core and BNCT filter.Two configurations were studied: Case $1$ corresponds to a water reflected configuration and it was the first critical configuration included in ICSBEP. Case $2$ corresponds to a graphite reflected configuration~\cite{bazzana}.

\section{Results}
\label{section:results}
\subsection{Effective multiplication factor}
\begin{sloppypar}
Results obtained for the effective multiplication factor calculated with MCNP and OpenMC(TD) are shown in Table~\ref{table:keff}, along with the experimental value when available, the difference between experimental and OpenMC(TD) calculated values, and the difference between MCNP and OpenMC(TD) calculated values. These criticality calculations were run using $100$ inactive and \numprint{2000} active batches of \numprint{200000} source neutrons, with the ENDF/B-VII.$1$ nuclear cross section libraries~\cite{chadwick}. 
By examining Table~\ref{table:keff} it can be seen that errors obtained for MCNP and OpenMC(TD) were less than $5$ and $6$~pcm, respectively, except for the RA-$6$ cases, where the simulation errors were $9$ and $10$~pcm. $k_\mathit{eff}$ values obtained from both simulations are within benchmark errors for $13$ of the $18$ cases and the differences between calculated values and benchmark errors were less than $0.3 \%$.
\end{sloppypar}

\begin{table*}[h!]
	\centering
	\footnotesize 
	{\def\arraystretch{1.2}\tabcolsep=8pt
		\begin{tabular}{@{}cccccccc@{}}
			\toprule
			Case & Case                   & $k_{\mathit{eff}}$   & $k_{\mathit{eff}}$  & $k_{\mathit{eff}}$  & $k_{\mathit{eff}}$ difference (pcm)      & $k_{\mathit{eff}}$ difference (pcm) & $k_{\mathit{eff}}$ difference (pcm) \\ 
		    \textnumero & name                 & exp.                 & MCNP                & OpenMC(TD)          &   Exp. - MCNP      & Exp. - OpenMC(TD)                   & MCNP - OpenMC(TD) \\ \midrule
			$1$ & Godiva                 & $1.00000(100)$         & $0.99977(5)$        & $0.99976(5)$       &  $23(100)$        & $24(100)$             & $1(7)$ \\
			$2$ & Jezebel-Pu             & $1.00000(200)$         & $0.99991(5)$        & $0.99985(5)$       &  $9(200)$        & $15(200)$             & $6(6)$ \\
			$3$ & Skidoo                 & $1.00000(100)$         & $0.99991(5)$        & $0.99991(5)$       &  $9(100)$        & $9(100)$              & $0(6)$ \\
			$4$ & Topsy                  & $1.00100(130)$         & $1.00156(5)$        & $1.00156(6)$       &  $-56(130)$        & $-56(130)$            & $0(8)$ \\
			$5$ & Popsy                  & $1.00000(300)$         & $1.00102(5)$        & $1.00096(6)$       &  $-102(300)$        & $-96(300)$            & $6(8)$ \\
			$6$ & Flattop$23$            & $1.00000(140)$         & $0.99880(5)$        & $0.99875(6)$       &  $120(140)$        & $125(140)$            & $5(8)$ \\
			$7$ & BigTen                 & $1.00490(80)$          & $1.00461(5)$        & $1.00462(5)$       &  $29(80)$        & $28(80)$              & $-1(6)$ \\
			$8$ & ZPR-U$9$               & $0.99540(240)$       & $0.99625(5)$        & $0.99623(5)$         &  $-85(240)$        & $-83(240)$            & $2(6)$ \\
			$9$ & SNEAK-$7$A            & $0.99980(350)$       & $1.00322(5)$        & $1.00310(6)$         &  $-342(350)$        & $-330(350)$           & $12(7)$ \\
			$10$ & SNEAK-$7$B            & $1.00010(400)$       & $0.99904(5)$        & $0.99943(5)$         &  $106(400)$      & $67(400)$             & $-39(6)$ \\
			$11$ & TCA                   & $1.00000(200)$       & $0.99999(5)$        & $1.00015(6)$         &  $1(200)$      & $-15(200)$            & $-16(7)$ \\
			$12$ & Stacy-$029$           & $0.99990(90)$        & $1.00141(5)$        & $1.00210(5)$         &  $-151(90)$      & $-220(90)$            & $-69(6)$ \\
			$13$ & Stacy-$033$           & $0.99990(90)$        & $0.99707(5)$        & $1.00002(5)$         &  $283(90)$      & $-12(90)$             & $-295(6)$ \\
			$14$ & Stacy-$046$           & $0.99990(100)$       & $1.00170(5)$        & $1.00226(5)$         &  $-180(100)$      & $-236(100)$           & $-56(6)$ \\
			$15$ & Stacy-$030$           & $0.99730(90)$        & $0.99705(5)$        & $0.99824(5)$         &  $25(90)$      & $-94(90)$             & $-119(6)$ \\
			$16$ & Stacy-$125$           & $0.99940(140)$       & $1.00498(5)$        & $1.00611(5)$         &  $-558(140)$      & $-671(140)$           & $-113(6)$ \\
			$17$ & RA-$6$ (Case $1$)     & $1.00160(140)$         & $1.00108(9)$        & $1.00132(10)$      &  $-191(250)$      & $28(140)$             & $-24(13)$ \\
			$18$ & RA-$6$ (Case $2$)     & $1.00260(250)$         & $1.00451(9)$        & $1.00589(10)$      &  $52(140)$      & $-329(250)$           & $-138(13)$\\ \bottomrule
		\end{tabular}                                                                                         
	}
	\caption{Results obtained for the effective multiplication factor. $k_\mathit{eff}$ was calculated using MCNP6 and OpenMC(TD). Both errors and differences are in pcm.}
	\label{table:keff}
\end{table*}

\subsection{Effective delayed neutron fraction}

\begin{table*}[h!]
	\centering
	\footnotesize 
	{\def\arraystretch{1.2}\tabcolsep=9pt
		\begin{tabular}{@{}ccccccccc@{}}
			\toprule
			Case &  Case                 & $\beta_{\mathit{eff}}$   & $\beta_{\mathit{eff}}$  & $\beta_{\mathit{eff}}$ & $\beta_{\mathit{eff}}$ difference & $\beta_{\mathit{eff}}$ difference & $\beta_{\mathit{eff}}$ difference \\ 
			\textnumero & Name                & Exp.                     & MCNP                    & OpenMC(TD)             &  Exp. - MCNP   & Exp. - OpenMC(TD)         & MCNP - OpenMC(TD) \\ \midrule
			$1$ & Godiva                 & $659(10)$                & $660(7)$                & $633(7)$         &  $-1(12)$     & $26(12)$              & $27(10)$ \\
			$2$ & Jezebel-Pu             & $194(10)$                & $190(4)$                & $181(7)$         &  $4(11)$     & $13(12)$              & $9(8)$ \\
			$3$ & Skidoo                 & $290(10)$                & $297(5)$                & $293(7)$         &  $-7(11)$     & $-3(12)$              & $4(9)$ \\
			$4$ & Topsy                  & $665(13)$                & $686(5)$                & $688(8)$         &   $-21(14)$    & $-23(16)$             & $-2(10)$ \\
			$5$ & Popsy                  & $276(7)$                 & $279(5)$                & $280(8)$         &   $-3(9)$    & $-4(11)$              & $-1(10)$ \\
			$6$ & Flattop$23$            & $360(9)$                 & $360(5)$                & $357(8)$         &   $0(10)$    & $3(12)$               & $3(10)$ \\
			$7$ & BigTen                 & $720(7)$                 & $711(7)$                & $696(7)$         &   $9(10)$    & $24(10)$              & $15(10)$ \\
			$8$ & ZPR-U$9$               & $725(17)$                & $722(7)$                & $700(7)$         &   $3(18)$    & $25(18)$              & $22(10)$ \\
			$9$ & SNEAK-$7$A            & $395(12)$                & $361(5)$                & $319(8)$         &    $34(13)$   & $76(14)$              & $42(9)$ \\
			$10$ & SNEAK-$7$B            & $429(13)$                & $411(6)$                & $374(7)$         &    $18(14)$   & $55(15)$              & $37(9)$ \\
			$11$ & TCA                   & $771(7)$                 & $765(6)$                & $766(10)$        &    $6(9)$   & $5(12)$               & $-1(12)$ \\
			$12$ & Stacy-$029$           & $-$                      & $730(6)$                & $713(8)$         &    $-$   & $-$                   & $17(10)$ \\
			$13$ & Stacy-$033$           & $-$                      & $719(6)$                & $718(8)$         &    $-$   & $-$                   & $1(10)$ \\
			$14$ & Stacy-$046$           & $-$                      & $718(6)$                & $709(8)$         &    $-$   & $-$                   & $9(9)$ \\
			$15$ & Stacy-$030$           & $-$                      & $732(6)$                & $736(8)$         &    $-$   & $-$                   & $-4(10)$ \\
			$16$ & Stacy-$125$           & $-$                      & $753(7)$                & $753(8)$         &    $-$   & $-$                   & $0(10)$ \\
			$17$ & RA-$6$ (Case $1$)     & $-$                      & $764(11)$               & $746(14)$        &    $-$   & $-$                   & $18(18)$ \\
			$18$ & RA-$6$ (Case $2$)     & $-$                      & $741(11)$               & $769(14)$        &    $-$   & $-$                   & $-28(18)$\\ \bottomrule
		\end{tabular}
	}
	\caption{Results obtained for the effective delayed neutron fraction. Effective delayed neutron fractions, statistical errors and differences are in pcm.}
	\label{table:betas}
\end{table*}
\begin{sloppypar}
Results obtained for the effective delayed neutron fraction calculated with MCNP and OpenMC(TD) are shown in Table~\ref{table:betas}, along with the experimental value when available, the difference between experimental and OpenMC(TD) calculated values, and the difference between MCNP and OpenMC(TD) calculated values. MCNP estimated $\beta_{\mathit{eff}}$ with adjoint-weighted tallies using the iterated fission probability method. OpenMC(TD) estimated $\beta_{\mathit{eff}}$ using the prompt method and the $k_p$ criticality calculations to compute $\beta_{\mathit{eff}}$ were run using $100$ inactive and \numprint{2000} active batches of \numprint{200000} source neutrons. All of these calculations used ENDF/B-VII.$1$ nuclear cross section libraries.

By examining Table~\ref{table:betas} it can be seen that differences between OpenMC(TD) calculated and benchmark values were less than $26$~pcm, except for SNEAK-$7$A ($76$~pcm) and SNEAK-$7$B ($55$~pcm). For MCNP differences between calculated and benchmark values were less than $34$~pcm. $\beta_\mathit{eff}$ values obtained from OpenMC(TD) are within benchmark errors for $4$ of the $12$ cases.

For Stacy-$029$, Stacy-$033$, Stacy-$046$, Stacy-$030$, Stacy-$125$ and both RA-$6$ cases there are no $\beta_{\mathit{eff}}$ experimental measurements available.
\end{sloppypar}
\subsection{Mean neutron generation time}
\begin{figure}[h!]
	\center
	{\includegraphics[width=1.0\columnwidth]{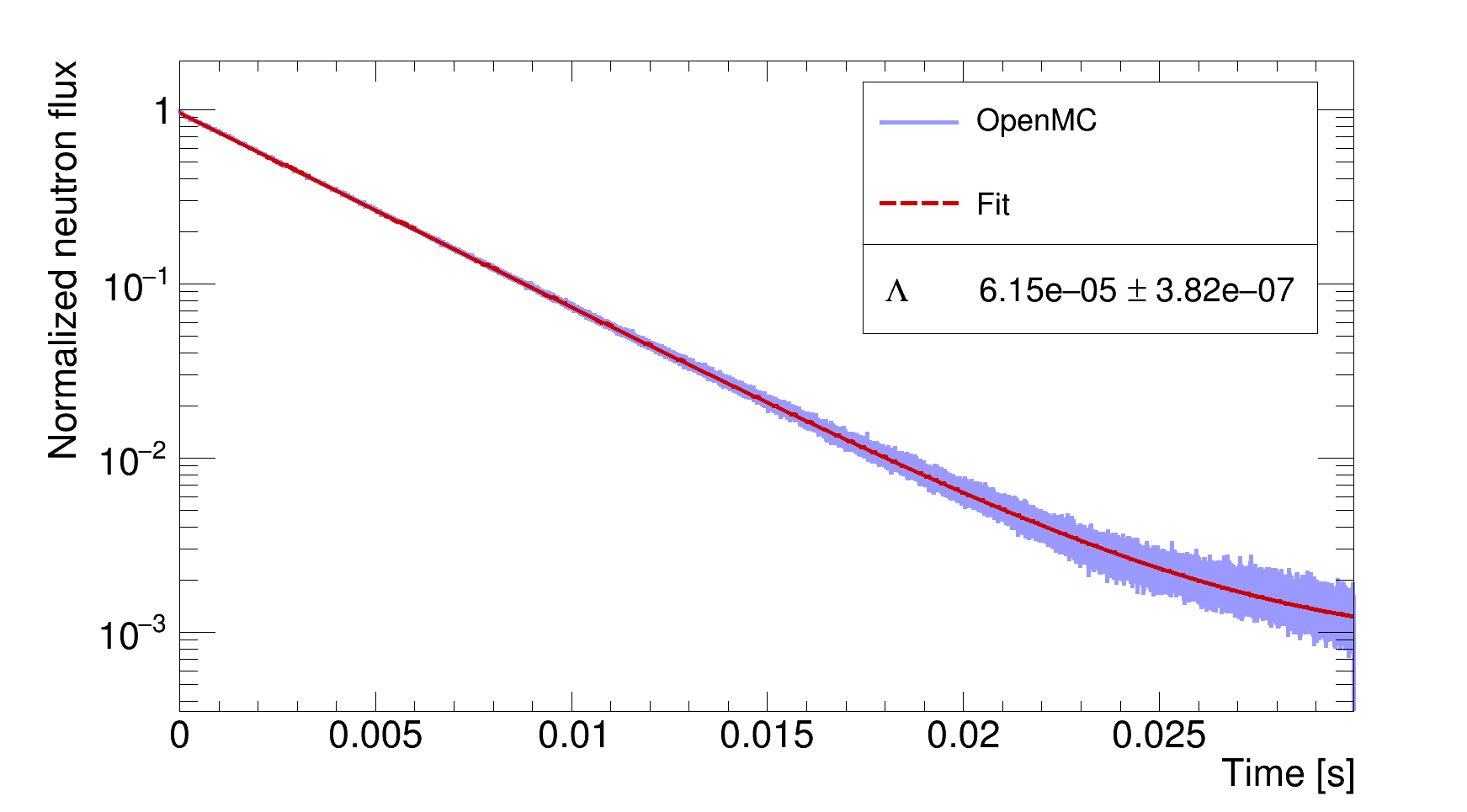}}
	\caption{Time evolution of the normalized neutron flux obtained using OpenMC(TD) for the benchmark Stacy-$029$. The time evolution of the neutron flux is shown in blue, while the fit obtained is shown in blue. $\Lambda$ value obtained from the fit is also shown.}
	\label{fig:stacy_config1_fit_log}	
\end{figure}

\begin{figure}[h!]
	\center
	{\includegraphics[width=1.0\columnwidth]{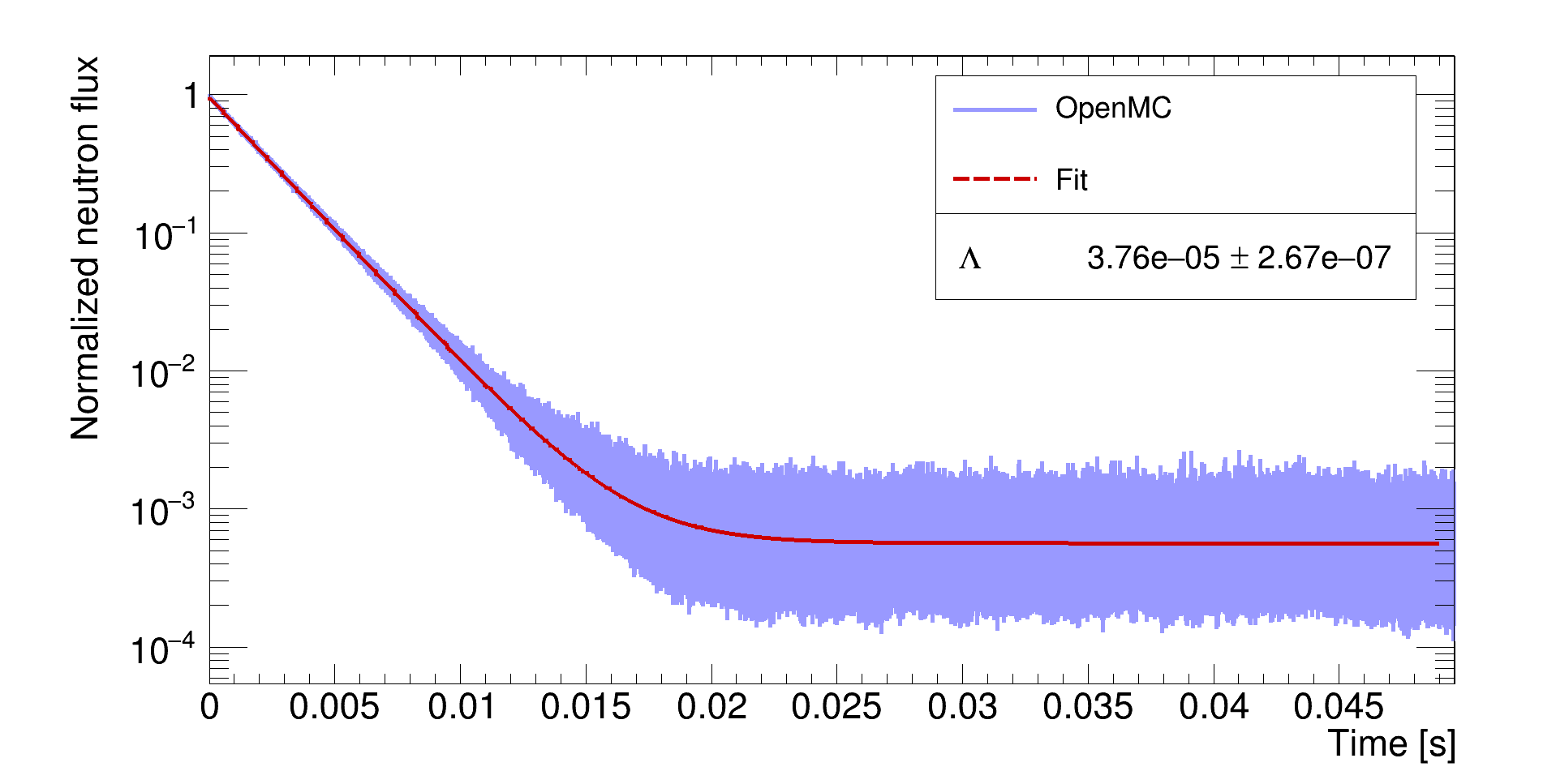}}
	\caption{Time evolution of the normalized neutron flux obtained using OpenMC(TD) for the benchmark RA-$6$ case $2$. The time evolution of the neutron flux is shown in blue, while the fit obtained is shown in blue. $\Lambda$ value obtained from the fit is also shown.}
	\label{fig:ra6_config1_fit_log}	
\end{figure}

\begin{sloppypar}
Neutron flux as a function of time calculated using OpenMC(TD) was obtained for the $18$ benchmark cases shown in Table~\ref{table:systems}. These fixed-source calculations were run using $100$ batches of \numprint{200000} source neutrons. As an example of results obtained, the neutron flux as a function of time for the Stacy-$030$ and RA-$6$ cases are shown in Fig.~\ref{fig:stacy_config1_fit_log} and Fig.~\ref{fig:ra6_config1_fit_log}, respectively. 
\end{sloppypar}

Results for the calculated mean neutron generation time obtained with MCNP and OpenMC(TD) are shown in Table~\ref{table:lambdas}. By examining results obtained it can be seen that the difference between MCNP and OpenMC(TD) are less than $7 \%$, except for Topsy, where the difference is of $8.1 \%$.

\begin{table*}[h!]
	\centering
	\footnotesize 
	{\def\arraystretch{1.2}\tabcolsep=12pt
		\begin{tabular}{@{}cccccc@{}}
			\toprule
			Case & Case                    & $\Lambda$    & $\Lambda$        & $\Lambda$ difference & Scaling \\ 
			Number & Name                  & MCNP         & OpenMC(TD)           & MCNP6 - OpenMC(TD)     &  Factor      \\ \midrule
			$1$ & Godiva                   & $5.689(8)$   & $5.699(95)$      & $-0.010(100)$  & $\times 10^{-9}$ \\
			$2$ & Jezebel-Pu               & $2.874(4)$   & $2.879(60)$      & $-0.005(60)$  & $\times 10^{-9}$ \\
			$3$ & Skidoo                   & $2.744(4)$   & $2.760(18)$      & $-0.016(18)$  & $\times 10^{-9}$ \\
			$4$ & Topsy                    & $1.791(4)$   & $1.936(6)$       & $-0.145(7)$ & $\times 10^{-8}$ \\
			$5$ & Popsy                    & $1.323(4)$   & $1.394(8)$       & $-0.071(9)$  & $\times 10^{-8}$ \\
			$6$ & Flattop$23$              & $1.258(4)$   & $1.330(7)$       & $-0.073(8)$   & $\times 10^{-8}$ \\
			$7$ & BigTen                   & $6.134(4)$   & $6.338(10)$      & $-0.204(11)$  & $\times 10^{-8}$ \\
			$8$ & ZPR-U$9$                 & $8.669(12)$  & $8.862(41)$      & $-0.193(43)$  & $\times 10^{-8}$ \\
			$9$ & SNEAK-$7$A              & $1.875(4)$   & $1.876(15)$      &  $-0.001(16)$  & $\times 10^{-7}$ \\
			$10$ & SNEAK-$7$B              & $1.728(3)$   & $1.748(15)$      & $-0.028(15)$  & $\times 10^{-7}$ \\
			$11$ & TCA                     & $3.522(3)$   & $3.742(32)$      & $-0.220(32)$   & $\times 10^{-5}$ \\
			$12$ & Stacy-$029$             & $5.962(5)$   & $6.147(38)$      & $-0.185(39)$       & $\times 10^{-5}$ \\
			$13$ & Stacy-$033$             & $6.315(6)$   & $6.456(38)$      & $-0.141(38)$       & $\times 10^{-5}$ \\
			$14$ & Stacy-$046$             & $6.716(5)$   & $6.993(39)$      & $-0.277(40)$       & $\times 10^{-5}$ \\
			$15$ & Stacy-$030$             & $5.853(4)$   & $5.860(32)$      & $-0.007(33)$       & $\times 10^{-5}$ \\
			$16$ & Stacy-$125$             & $4.851(4)$   & $5.111(33)$      & $-0.260(33)$       & $\times 10^{-5}$ \\
			$17$ & RA-$6$ (Case $1$)       & $4.177(10)$  & $4.296(29)$     &  $-0.119(31)$       & $\times 10^{-5}$ \\
			$18$ & RA-$6$ (Case $2$)       & $3.637(7)$   & $3.757(27)$     &  $-0.120(28)$       & $\times 10^{-5}$\\ \bottomrule
		\end{tabular}
	}
	\caption{Results obtained for the prompt generation times. Effective prompt generation times, statistical errors and differences are in seconds and the scaling factor is given in the last column.}
	\label{table:lambdas}
\end{table*}

In systems from Table~\ref{table:systems}, $\Lambda$ is not reported explicitly, nevertheless, $\alpha_R=\beta_{\mathit{eff}} /\Lambda$~\cite{LA-UR-11-04409} is reported for some of the benchmarks and its results are compared to simulations in Table~\ref{table:alphas}. It can be seen that for OpenMC(TD) the difference between benchmark and calculated values is less than $6.56 \%$, while for MCNP this difference is less than $4.31 \%$ except for Flattop$23$ case where it is $6.64 \%$.
\begin{table*}[h!]
	\centering
	\footnotesize 
	{\def\arraystretch{1.2}\tabcolsep=8pt
		\begin{tabular}{@{}ccccccccc@{}}
			\toprule
			Case   & Case              & $\alpha_R$    &   $\alpha_R$     & $\alpha_R$    & $\alpha_R$ difference  & $\alpha_R$ difference & $\alpha_R$ difference & Scaling \\ 
			Number & Name              & Exp.          &   MCNP           & OpenMC(TD)    & Exp. - MCNP      & Exp. - OpenMC(TD) & MCNP - OpenMC(TD)    & Factor       \\ \midrule
			$1$    & Godiva            & $-1.11(2)$     & $-1.16(2)$      & $-1.11(2)$    &  $0.05(3)$  & $0.00(3)$             &  $0.05(3)$     & $\times 10^{6}$ \\
			$2$    & Jezebel-Pu        & $-6.4(1)$      & $-6.60(13)$     & $-6.28(28)$   & $0.20(13)$  & $-0.12(29)$           & $-0.32(31)$   & $\times 10^{5}$ \\
			$6$    & Flattop$23$       & $-2.67(5)$     & $-2.86(4)$      & $-2.68(16)$   & $0.19(6)$  & $0.01(17)$            & $-0.18(16)$   & $\times 10^{5}$ \\
			$7$    & BigTen            & $-1.170(10)$   & $-1.159(12)$    & $-1.098(21)$  & $-0.011(16)$  & $-0.072(23)$          & $0.061(24)$      & $\times 10^{5}$ \\
			$12$   & Stacy-$029$       & $-1.227(41)$   & $-1.224(11)$    & $-1.154(14)$  & $-0.003(42)$  & $-0.073(43)$          & $0.070(18)$   & $\times 10^{2}$ \\
			$13$   & Stacy-$033$       & $-1.167(39)$   & $-1.138(9)$     & $-1.101(12)$  & $-0.029(40)$  & $-0.066(41)$          & $-0.037(15)$  & $\times 10^{2}$ \\
			$14$   & Stacy-$046$       & $-1.062(37)$   & $-1.069(9)$     & $-1.005(15)$  & $0.007(38)$  & $-0.057(40)$          & $-0.064(17)$  & $\times 10^{2}$ \\
			$15$   & Stacy-$030$       & $-1.268(29)$   & $-1.251(11)$    & $-1.197(14)$  & $-0.017(31)$  & $-0.071(32)$          & $-0.054(18)$  & $\times 10^{2}$ \\
			$16$   & Stacy-$125$       & $-1.528(26)$   & $-1.553(14)$    & $-1.464(17)$  & $0.025(30)$  & $-0.064(31)$          & $-0.089(22)$  & $\times 10^{2}$ \\
			$18$   & RA-$6$ (Case $2$) & $-1.808(9)$    & $-1.829(27)$    & $-1.724(27)$  & $0.021(28)$  & $-0.084(28)$          & $-0.105(38)$  & $\times 10^{2}$ \\ \bottomrule
		\end{tabular}
	}
	\caption{Results obtained for the measured Rossi-$\alpha$. Effective delayed neutron fractions, statistical errors and differences are in pcm.}
	\label{table:alphas}
\end{table*}

\section{Summary and conclusions}
\label{section:summary}
This work presented the methodology developed to include time dependence for the open source Monte Carlo code OpenMC. The modified code, OpenMC(TD), was used to calculate the effective delayed neutron fraction $\beta_{\mathit{eff}}$ using the prompt method. The mean neutron generation time $\Lambda$ was estimated by scoring the decay of the neutron population after using the pulsed method.  
MCNP estimates $\beta_{\mathit{eff}}$, $\Lambda$, and $\alpha_R$ using adjoint-weighted tallies in a criticality calculation. In OpenMC, $\beta_{\mathit{eff}}$ was obtained using the prompt method approximation, while $\Lambda$ and $\alpha_R$ were calculated using OpenMC(TD) new time-dependence capabilities, through a fit, as it was shown in Fig~\ref{fig:stacy_config1_fit_log} and Fig.\ref{fig:ra6_config1_fit_log}. It is important to notice that the pulsed method is an approximation that requires a slightly subcritical system to observe the decay of the neutron flux in a fixed-source calculation. Nevertheless, simulation results obtained with OpenMC(TD) are as good as the results obtained using MCNP.


\section{Acknowledgments}

J.Romero-Barrientos acknowledges support from Programa Nacional de Becas de Postgrado under grant $21151413$, and from FONDECYT Regular Project $1171467$.


\bibliography{mybibfile}

\begin{thebibliography}{10}
\expandafter\ifx\csname url\endcsname\relax
  \def\url#1{\texttt{#1}}\fi
\expandafter\ifx\csname urlprefix\endcsname\relax\def\urlprefix{URL }\fi
\expandafter\ifx\csname href\endcsname\relax
  \def\href#1#2{#2} \def\path#1{#1}\fi

\bibitem{Goorley2012}
T.~Goorley, M.~James, T.~Booth, F.~Brown, J.~Bull, L.~J. Cox, J.~Durkee,
  J.~Elson, M.~Fensin, R.~A. Forster, J.~Hendricks, H.~G. Hughes, R.~Johns,
  B.~Kiedrowski, R.~Martz, S.~Mashnik, G.~McKinney, D.~Pelowitz, R.~Prael,
  J.~Sweezy, L.~Waters, T.~Wilcox, T.~Zukaitis, {Initial MCNP6 Release
  Overview}, Nucl. Technol. 180~(3) (2012) 298--315.
\newblock \href {http://dx.doi.org/10.13182/NT11-135}
  {\path{doi:10.13182/NT11-135}}.

\bibitem{Romano2015}
P.~K. Romano, N.~E. Horelik, B.~R. Herman, A.~G. Nelson, B.~Forget, K.~Smith,
  {OpenMC: A state-of-the-art Monte Carlo code for research and development},
  Annals of Nuclear Energy 82 (2015) 90 -- 97, {Joint International Conference
  on Supercomputing in Nuclear Applications and Monte Carlo 2013}.
\newblock \href
  {http://dx.doi.org/https://doi.org/10.1016/j.anucene.2014.07.048}
  {\path{doi:https://doi.org/10.1016/j.anucene.2014.07.048}}.

\bibitem{icsbep}
J.~B. Briggs, L.~Scott, A.~Nouri, The international criticality safety
  benchmark evaluation project, Nuclear Science and Engineering 145~(1) (2003)
  1--10.
\newblock \href {http://dx.doi.org/10.13182/NSE03-14}
  {\path{doi:10.13182/NSE03-14}}.

\bibitem{leppanen}
J.~Leppänen, M.~Aufiero, E.~Fridman, R.~Rachamin, S.~{van der Marck},
  {Calculation of effective point kinetics parameters in the Serpent 2 Monte
  Carlo code}, Annals of Nuclear Energy 65 (2014) 272--279.
\newblock \href
  {http://dx.doi.org/https://doi.org/10.1016/j.anucene.2013.10.032}
  {\path{doi:https://doi.org/10.1016/j.anucene.2013.10.032}}.

\bibitem{Koranne2011}
S.~Koranne, Hierarchical Data Format 5 : HDF5, Springer US, Boston, MA, 2011,
  pp. 191--200.
\newblock \href {http://dx.doi.org/10.1007/978-1-4419-7719-9_10}
  {\path{doi:10.1007/978-1-4419-7719-9_10}}.

\bibitem{osti_1338791}
R.~Macfarlane, D.~W. Muir, R.~M. Boicourt, A.~C. Kahler, III, J.~L. Conlin,
  {The NJOY Nuclear Data Processing System, Version 2016}, Technical Report Los
  Alamos National Laboratory~(TRN: US1701456).
\newblock \href {http://dx.doi.org/10.2172/1338791}
  {\path{doi:10.2172/1338791}}.

\bibitem{khurrum}
K.~S. Chaudri, S.~M. Mirza, {Burnup dependent Monte Carlo neutron physics
  calculations of IAEA MTR benchmark}, Progress in Nuclear Energy 81 (2015)
  43--52.
\newblock \href
  {http://dx.doi.org/https://doi.org/10.1016/j.pnucene.2014.12.018}
  {\path{doi:https://doi.org/10.1016/j.pnucene.2014.12.018}}.

\bibitem{CHEN2017264}
J.~Chen, L.~Cao, C.~Zhao, Z.~Liu, {Development of Subchannel Code SUBSC for
  high-fidelity multi-physics coupling application}, Energy Procedia 127 (2017)
  264--274, international Youth Nuclear Congress 2016, IYNC2016, 24-30 July
  2016, Hangzhou, China.
\newblock \href
  {http://dx.doi.org/https://doi.org/10.1016/j.egypro.2017.08.121}
  {\path{doi:https://doi.org/10.1016/j.egypro.2017.08.121}}.

\bibitem{LI2018329}
J.-Y. Li, L.~Gu, H.-S. Xu, N.~Korepanova, R.~Yu, Y.-L. Zhu, C.-P. Qin, {CAD
  modeling study on FLUKA and OpenMC for accelerator driven system simulation},
  Annals of Nuclear Energy 114 (2018) 329--341.
\newblock \href
  {http://dx.doi.org/https://doi.org/10.1016/j.anucene.2017.12.050}
  {\path{doi:https://doi.org/10.1016/j.anucene.2017.12.050}}.

\bibitem{variansyah}
I.~Variansyah, B.~R. Betzler, W.~R. Martin, {Multigroup Constant Calculation
  with Static $\alpha$-Eigenvalue Monte Carlo for Time-Dependent Neutron
  Transport Simulations}, Nuclear Science and Engineering 194~(11) (2020)
  1025--1043.
\newblock \href {http://dx.doi.org/10.1080/00295639.2020.1743578}
  {\path{doi:10.1080/00295639.2020.1743578}}.

\bibitem{walsh}
J.~A. Walsh, B.~Forget, K.~S. Smith, F.~B. Brown, {On-the-fly Doppler
  broadening of unresolved resonance region cross sections}, Progress in
  Nuclear Energy 101 (2017) 444--460, special Issue on the Physics of Reactors
  International Conference PHYSOR 2016: Unifying Theory and Experiments in the
  21st Century.
\newblock \href
  {http://dx.doi.org/https://doi.org/10.1016/j.pnucene.2017.05.032}
  {\path{doi:https://doi.org/10.1016/j.pnucene.2017.05.032}}.

\bibitem{walsh_j}
J.~A. Walsh, P.~K. Romano, B.~Forget, K.~S. Smith, {Optimizations of the energy
  grid search algorithm in continuous-energy Monte Carlo particle transport
  codes}, Computer Physics Communications 196 (2015) 134--142.
\newblock \href {http://dx.doi.org/https://doi.org/10.1016/j.cpc.2015.05.025}
  {\path{doi:https://doi.org/10.1016/j.cpc.2015.05.025}}.

\bibitem{forget}
P.~K. Romano, A.~R. Siegel, B.~Forget, K.~Smith, {Data decomposition of Monte
  Carlo particle transport simulations via tally servers}, Journal of
  Computational Physics 252 (2013) 20--36.
\newblock \href {http://dx.doi.org/https://doi.org/10.1016/j.jcp.2013.06.011}
  {\path{doi:https://doi.org/10.1016/j.jcp.2013.06.011}}.

\bibitem{binhang}
B.~Zhang, X.~Yuan, Y.~Zhang, H.~Tang, L.~Cao, {Development of a versatile
  depletion code AMAC}, Annals of Nuclear Energy 143 (2020) 107446.
\newblock \href
  {http://dx.doi.org/https://doi.org/10.1016/j.anucene.2020.107446}
  {\path{doi:https://doi.org/10.1016/j.anucene.2020.107446}}.

\bibitem{xingjie}
X.~Peng, J.~Liang, B.~Forget, K.~Smith, {Calculation of adjoint-weighted
  reactor kinetics parameters in OpenMC}, Annals of Nuclear Energy 128 (2019)
  231--235.
\newblock \href
  {http://dx.doi.org/https://doi.org/10.1016/j.anucene.2019.01.007}
  {\path{doi:https://doi.org/10.1016/j.anucene.2019.01.007}}.

\bibitem{arxiv-moi}
J.~Romero-Barrientos, Time-dependent monte carlo in fissile systems with
  beta-delayed neutron precursors, Ph.D. thesis (2021).
\newblock \href {http://arxiv.org/abs/2101.09338} {\path{arXiv:2101.09338}}.

\bibitem{LA-UR-10-01700}
B.~C. Kiedrowski, {Theory, Interface, Verification, Validation, and Performance
  of the Adjoint-Weighted Point Reactor Kinetics Parameter Calculations in
  MCNP}, {LA-UR-10-01700 Los Alamos National Laboratory}.

\bibitem{Meulekamp2006}
R.~K. Meulekamp, S.~C. van~der Marck, {Calculating the Effective Delayed
  Neutron Fraction with Monte Carlo}, Nuclear Science and Engineering 152~(2)
  (2006) 142--148.
\newblock \href {http://dx.doi.org/10.13182/NSE03-107}
  {\path{doi:10.13182/NSE03-107}}.

\bibitem{simmons}
B.~E. Simmons, J.~S. King, {A Pulsed Neutron Technique for Reactivity
  Determination}, Nuclear Science and Engineering 3~(5) (1958) 595--608.
\newblock \href {http://dx.doi.org/10.13182/NSE3-595-608}
  {\path{doi:10.13182/NSE3-595-608}}.

\bibitem{snoj}
L.~Snoj, A.~Kavčič, G.~Žerovnik, M.~Ravnik, {Calculation of kinetic
  parameters for mixed TRIGA cores with Monte Carlo}, Annals of Nuclear Energy
  37~(2) (2010) 223--229.
\newblock \href
  {http://dx.doi.org/https://doi.org/10.1016/j.anucene.2009.10.020}
  {\path{doi:https://doi.org/10.1016/j.anucene.2009.10.020}}.

\bibitem{bell}
G.~I. Bell, S.~Glasstone, {Nuclear reactor theory}, Van Nostrand Reinhold,
  1970.

\bibitem{bazzana}
S.~Bazzana, {Desarrollo, an\'alisis y evaluaci\'on de experimentos
  neutr\'onicos en el RA-$6$}, Master's thesis, Instituto Balseiro, Universidad
  Nacional de Cuyo, Comisi\'on Nacional de Energ\'ia At\'omica, Argentina, San
  Carlos de Bariloche, Argentina (2012).

\bibitem{chadwick}
M.~Chadwick, M.~Herman, P.~Obložinský, M.~Dunn, Y.~Danon, A.~Kahler,
  D.~Smith, B.~Pritychenko, G.~Arbanas, R.~Arcilla, R.~Brewer, D.~Brown,
  R.~Capote, A.~Carlson, Y.~Cho, H.~Derrien, K.~Guber, G.~Hale, S.~Hoblit,
  S.~Holloway, T.~Johnson, T.~Kawano, B.~Kiedrowski, H.~Kim, S.~Kunieda,
  N.~Larson, L.~Leal, J.~Lestone, R.~Little, E.~McCutchan, R.~MacFarlane,
  M.~MacInnes, C.~Mattoon, R.~McKnight, S.~Mughabghab, G.~Nobre, G.~Palmiotti,
  A.~Palumbo, M.~Pigni, V.~Pronyaev, R.~Sayer, A.~Sonzogni, N.~Summers,
  P.~Talou, I.~Thompson, A.~Trkov, R.~Vogt, S.~{van der Marck}, A.~Wallner,
  M.~White, D.~Wiarda, P.~Young, {ENDF/B-VII.1 Nuclear Data for Science and
  Technology: Cross Sections, Covariances, Fission Product Yields and Decay
  Data}, Nuclear Data Sheets 112~(12) (2011) 2887--2996, special Issue on
  ENDF/B-VII.1 Library.
\newblock \href {http://dx.doi.org/https://doi.org/10.1016/j.nds.2011.11.002}
  {\path{doi:https://doi.org/10.1016/j.nds.2011.11.002}}.

\bibitem{LA-UR-11-04409}
R.~D. Mosteller, B.~C. Kiedrowski, {The Rossi alpha validation suite for MCNP},
  {LA-UR-11-04409 Los Alamos National Laboratory}.

\end{thebibliography}
\balance


%

\end{document}